\documentclass[12pt]{article}
\usepackage{amsmath}
\usepackage{amssymb}
\begin{document}
\vskip 2cm
\begin{center}
{\sf {\Large Supervariable approach to nilpotent symmetries of  a couple of $\mathcal{N }= 2$ 
supersymmetric quantum mechanical models}}

\vskip 3.0cm

{\sf S. Krishna$^{(a)}$, A. Shukla$^{(a)}$, R. P. Malik$^{(a,b)}$}\\
$^{(a)}$ {\it Physics Department, Centre of Advanced Studies,}\\
{\it Banaras Hindu University, Varanasi - 221 005, (U.P.), India}\\

\vskip 0.1cm

{\bf and}\\

\vskip 0.1cm

$^{(b)}$ {\it DST Centre for Interdisciplinary Mathematical Sciences,}\\
{\it Faculty of Science, Banaras Hindu University, Varanasi - 221 005, India}\\
{\small {\sf {e-mails: skrishna.bhu@gmail.com; ashukla038@gmail.com;  rpmalik1995@gmail.com}}}

\end{center}

\vskip 2cm

\noindent
{\bf Abstract:} 
We derive the on-shell as well as off-shell nilpotent supersymmetric (SUSY) symmetry 
transformations for the  $\mathcal {N} = 2$ SUSY quantum 
mechanical model of a one (0+1)-dimensional (1D) free SUSY particle by exploiting the SUSY 
invariant restrictions  (SUSYIRs)
on the (anti-)chiral supervariables of the SUSY theory that is defined on a (1, 2)-dimensional 
supermanifold (parametrized by a bosonic variable $t$ and a pair of Grassmannian variables
 $\theta$ and $\bar\theta$ with $\theta^2 = \bar\theta^2 = 0,\, 
\theta\bar\theta + \bar\theta\theta = 0$). Within the framework of our novel approach, 
we express the Lagrangian and conserved SUSY charges in terms of the (anti-)chiral supervariables 
to demonstrate the SUSY invariance of the Lagrangian as well as the nilpotency of the SUSY conserved charges
 in a simple manner. Our approach has the potential to be generalized to the description of 
 other $\mathcal {N} = 2$ SUSY
quantum mechanical systems  with physically interesting potential functions.
To corroborate the above assertion, we apply our method to derive the
 $\mathcal{N} = 2$ continuous and nilpotent SUSY transformations
for one of the simplest {\it interacting} SUSY system of a 1D harmonic oscillator.

\vskip 0.8cm
\noindent
PACS numbers: 11.30.Pb; 03.65.-w; 11.30.-j

\vskip 0.5cm
\noindent
{\it Keywords}: $\mathcal{N }= 2$ SUSY quantum mechanics; $\mathcal{N }= 2$ SUSY free particle and harmonic 
oscillator; on-shell and off-shell nilpotent symmetries; SUSY invariant restrictions; $\mathcal{N }= 2$ 
SUSY algebra and its cohomological interpretation

\newpage

\noindent

\section{Introduction}
\label{intro}
One of the most elegant, intuitive and geometrically rich approaches to derive the ``quantum'' 
symmetries corresponding to the 
``classical'' gauge and/or reparameteriztion symmetries is the superfield formalism [1-8]. In particular,
Bonora-Tonin (BT) superfield approach [4,5] is very suitable for the derivation of the off-shell
nilpotent and absolutely anticommuting Becchi-Rouet-Stora-Tyutin (BRST) and anti-BRST symmetries for a given 
$p$-form ($p = 1, 2, 3,...$) (non-)Abelian
gauge theory in any arbitrary dimension of spacetime where the celebrated horizontality condition (HC)
plays a very decisive role. 
The HC leads to the derivation of ``quantum'' gauge [i.e. (anti-)BRST] symmetries for the bosonic gauge
and corresponding fermionic (anti-)ghost fields of the 
(anti-) BRST invariant gauge theory. This statement is true in any arbitrary dimension of spacetime.

In a set of papers (see, e.g. [9-12]), the {\it augmented} version of BT-formalism 
has been developed where, in addition to the HC, 
the {\it gauge invariant} restrictions (GIRs) on the superfields
have been also imposed   for the derivation of the ``quantum'' gauge [i.e. (anti-) BRST]
symmetries for the matter fields, in addition to the (anti-)BRST symmetries
for the gauge and corresponding (anti-)ghost fields (which emerge from the HC)
for a given $p$-form {\it interacting} (non-)Abelian gauge theory. As it has turned out, 
the nilpotent (anti-) BRST symmetries and their geometrical interpretations  derived from the
HC and GIRs, have been found to be consistent with one-another thereby leading to the derivation
of   full set of {\it proper} (i.e. nilpotent and absolutely anticommuting) (anti-)BRST symmetries for a 
given $p$-form interacting gauge theory.

The above superfield formalisms [1-12] have, however,  {\it not} yet been exploited in the context of
supersymmetric (SUSY) theories where the nilpotent symmetries are {\it also} found to exist. The purpose of our 
present investigation is to exploit the key {\it ideas} of the augmented version of superfield
formalism [9-12] (without any use of the HC) to derive the on-shell nilpotent SUSY symmetries of
one of the simplest $\mathcal{N} = 2$ SUSY quantum mechanical models of the free SUSY particle\footnote{We also 
briefly mention about the derivation of  $\mathcal{N} = 2$ SUSY symmetries  for an {\it interacting}
1D model of SUSY harmonic oscillator in our Appendix A and derive the {\it off-shell} nilpotent
$\mathcal{N} = 2$ SUSY symmetries for the free SUSY particle in our Appendix B.}. 
We accomplish our goals by taking the help of chiral and anti-chiral supervariables and imposing SUSY
{\it invariant} restrictions  (SUSYIRs) on them. Our method of derivation of  SUSY 
transformations, to the best of our knowledge, 
has never been exploited in the context of $\mathcal {N} = 2$ 
SUSY quantum mechanical (and/or field theoretic) examples as far as the derivation of symmetries is concerned.

The main motivating factors behind our present investigation are as follows. First, it is very challenging to find out
an alternative to the usual mathematical method of deriving the SUSY symmetries for a SUSY quantum mechanical model. 
In our present investigation, we accomplish this goal by exploiting the SUSY invariance. Second, 
our approach is {\it physically} more
appealing because we exploit the SUSY invariance to put restrictions on the (anti-) chiral supervariables.
Third, to prove the generality of our approach, we discuss one of the simplest {\it interacting} $\mathcal{N} = 2$ SUSY system 
of a harmonic oscillator in our Appendix A.
Finally, our present attempt is our modest {\it first} step in the direction of developing a general rule 
to derive the SUSY symmetry transformations for the 
$\mathcal{N} = 2$ SUSY systems of physical interest.

The contents of our present investigation are organized as follows. In our Sec. 2, 
we exploit the bare essentials of $\mathcal{N }= 2$ superspace formulation to derive the 
action (and/or Lagrangian) for our present model of $\mathcal{N }= 2$ SUSY free particle. 
Our Sec. 3 is devoted to the derivation of SUSY transformations generated by one of the 
two SUSY charges by using the anti-chiral supervariables. We derive the other SUSY 
transformations in Sec. 4 by exploiting the chiral supervariables. 
Our  Sec. 5 contains the derivation of  standard $\mathcal{N }= 2$ SUSY algebra and 
we provide its interpretation in the language of
cohomological operators of  differential geometry.
Finally, we make some concluding remarks in Sec. 6 and say a few words about the reasons behind 
our choice of (anti-)chiral supervariables in the context of our present discussion.

In our Appendix A, we apply our method to derive the off-shell as well as 
on-shell  $\mathcal{N} = 2$ SUSY continuous symmetry  transformations for one of the 
simplest {\it interacting}  SUSY system  of a one (0 + 1)-dimensional harmonic oscillator.
Our Appendix B is devoted to the derivation of $\mathcal{N} = 2$ off-shell nilpotent 
($s^2_1 = s_2^2 = 0$)  SUSY symmetry  transformations for the free SUSY particle.

{\bf \it{General notations and conventions:}} Throughout the whole  body our of text, 
we shall denote the fermionic ($s^2_1 = s_2^2 = 0$) $\mathcal{N} = 2$
SUSY transformations by $s_1$ and $s_2$ and shall adopt the convention of the left 
derivative w.r.t. the fermionic variables. We shall
use the notations $\big(x(t), \psi(t), \bar\psi(t)\big)$ for the bosonic variable ($x$) 
and a pair of fermionic variables $\psi$ and $\bar\psi$
for the $\mathcal{N} = 2$ SUSY QM theory. The corresponding supervariables would be denoted by $X, \Psi, \bar\Psi$
which would  be defined on the (anti-)chiral super-submanifolds. The auxiliary (super)variables would be represented 
by $(\tilde A)A$.

\section {Preliminaries: $\mathcal{N }= 2$ superspace approach to derive the action integral 
and $\mathcal{N }= 2$ SUSY symmetries  for the free SUSY particle}

We begin with the supervariable [$X(Z) \equiv X(t, \theta, \bar\theta)$] where the $\mathcal{N }= 2$ 
superspace coordinates $Z^M = (t, \theta, \bar\theta)$ are parameterized by the bosonic evolution  
parameter $t$ and a pair of Grassmannian variables $\theta$ and $\bar\theta$ 
(with $\theta^2 = \bar\theta^2 = 0, \, \theta\bar\theta + \bar\theta \theta = 0$). 
The above supervariable can be expanded along the Grassmannian directions $\theta$ 
and $\bar\theta$ of the (1, 2)-dimensional supermanifold (on which our present theory
is considered) as follows (see, e.g. [13,14])
\begin{eqnarray}
X(t,\theta,\bar\theta) = x(t) + i \,\theta\, \bar\psi(t) + i\, \bar\theta\, \psi(t) + \theta\,\bar\theta\, A(t),
\end{eqnarray}
where, on the r.h.s., the component variables    ($x, A$) are bosonic and ($\psi,\,\bar\psi$)  
are fermionic ($\psi^2 = \bar\psi^2 = 0,\, \psi \bar\psi + \bar\psi \psi = 0$) at the {\it classical} 
level. The basic dynamical variables of the $\mathcal{N }= 2$ SUSY free particle
are $x(t), \psi(t)$ and $\bar\psi(t)$, in terms of which, the Lagrangian function is defined. The latter
 includes the corresponding generalized ``velocities" (i.e. $\dot x(t), \dot\psi(t), \dot{\bar\psi}(t)$), too. 
The auxiliary  variable $A(t)$ does {\it not} play any role in the description of the 
{\it free} $\mathcal{N }= 2$ SUSY
particle because it is connected with the {\it potential function} of a given physical system. 
Thus, for our further discussions (on the free SUSY  particle within the supervariable approach), 
we set $A = 0$, right from the beginning, in the above expansion (1) for the supervariable $X(t,\theta,\bar\theta)$.

The action integral for the $\mathcal{N }= 2$ SUSY {\it free} particle can be written as [13-15]
\begin{eqnarray}
S = \int dt\, \int d\theta\, \int d\bar\theta\,\, {\cal D}X(t,\theta,\bar\theta)\,\bar {\cal D} 
X(t,\theta,\bar\theta),
\end{eqnarray}
where ${\cal D}$ and $\bar {\cal D}$ are the super covariant derivatives
\begin{eqnarray}
 &&{\cal D}  = \frac{\partial}{\partial\bar\theta} - i\,\theta\, \frac{\partial}{\partial t} 
\equiv \partial_{\bar\theta} - i\,\theta\,\partial_t,\qquad
 \bar {\cal D} = \frac{\partial}{\partial\theta} - i\,\bar\theta \,\frac{\partial}{\partial t}
\equiv \partial_{\theta} - i\,\bar\theta\,\partial_t.
\end{eqnarray}
The  superspace derivatives $\partial_M = {\partial}/{\partial Z^M} 
\equiv (\partial_t, \partial_{\theta}, \partial_{\bar\theta})$
are the generators of the shift transformations along the superspace coordinates as: 
\begin{eqnarray}
&& t\longrightarrow t^{\prime} = t + i\,(\varepsilon \,\bar\theta + \bar\varepsilon \,\theta), \qquad
\theta\longrightarrow \theta^{\prime} = \theta + \varepsilon,\qquad\bar\theta
\longrightarrow \bar\theta^{\prime} = \bar\theta + \bar\varepsilon,
\end{eqnarray} 
where $\varepsilon$ and $\bar\varepsilon$ are the infinitesimal time-independent shift parameters along the Grassmannian  
directions $\theta $ and $\bar\theta $ of 
the (1, 2)-dimensional supermanifold. As a consequence, these parameters are {\it also} fermionic 
in nature ($\varepsilon^2= \bar\varepsilon^2 =0,
 \varepsilon\,\bar\varepsilon + \bar\varepsilon\,\varepsilon = 0$).

Substitutions of $X(t, \theta, \bar\theta)$ from (1) (with $A = 0$) and the 
operations of the super covariant  derivatives (3) on them, finally, lead to the derivation of 
the action integral ($S = \int dt\, L_0$) for the $\mathcal{N }= 2$  SUSY free particle as follows:
\begin{eqnarray}
S = \int dt\,\Big[\frac{1}{2}\,\dot x^2 - \frac{i}{2}\,(\dot{\bar\psi}\,\psi 
- \bar\psi\,\dot\psi) \Big] \equiv  \int dt \, L_0,
\end{eqnarray}
where we  have already performed the Grassmannian integrations and, 
for the sake of brevity, we have chosen  the mass ($m$) of the 
$\mathcal{N }= 2$  SUSY free particle to be one
(i.e. $m = 1$). Ultimately, we obtain the Lagrangian ($L_0$) for the free particle as
 \begin{eqnarray}
L_0 = \frac{1}{2}\,\dot x^2 + i\, \bar\psi\,\dot\psi,\qquad\qquad\qquad (m = 1),
\end{eqnarray}
where $\dot x = (dx/dt), \, \dot\psi = (d\psi/dt)$ and we have dropped a total time derivative term.
It is elementary to check that the Euler-Lagrange (EL) equations of motion are:
$\ddot x = 0,\, \dot{\bar\psi} = 0,\, \dot\psi = 0$. These EL equations of motion ensure  that 
there  is {\it no} potential function (and/or force) for the description of  our present model 
(which is nothing but the $\mathcal {N} = 2$ SUSY {\it free} particle). 
The analogue of the Lagrangian (6) can be derived from the general $\mathcal{N} = 2$ superspace approach
for any arbitrary potential function (see, e.g. [15]). For instance, the Lagrangian for the $\mathcal {N} = 2$
SUSY harmonic oscillator (see Appendix A) is a special case of  $\mathcal{N} = 2$ SUSY theory 
for the general potential function (see, e.g. [13-15] for details) where $A(t)$ is chosen in a particular fashion
(i.e. $A = \omega\, x$).

Two SUSY transformations ($\delta_1$ and $\delta_2$) can be computed from the standard superspace formula for the 
$\mathcal{N} = 2$ SUSY quantum mechanical theory. These transformations are 
\begin{eqnarray}
&&(\delta_1+ \delta_2)\,X(t,\theta,\bar\theta) = (\delta_1+ \delta_2)\, x(t) + i \,\theta\,(\delta_1+ \delta_2)\, \bar\psi(t) + i\, \bar\theta\, (\delta_1+ \delta_2)\,\psi(t)  \nonumber\\  
&& + \theta\,\bar\theta\,(\delta_1+ \delta_2)\, A(t)
\equiv (\varepsilon \bar Q + \bar\varepsilon Q)\,X(t,\theta,\bar\theta),
\end{eqnarray}
where $Q$ and $\bar Q$ are the $\mathcal{N} = 2$ SUSY (fermionic) charges (with $Q^2 = {\bar Q}^2 = 0$) that are defined, 
in their  operator form, as follows: 
\begin{eqnarray}
Q = \partial_{\bar\theta} + i \theta\, \partial_t, \qquad
\qquad\quad \bar Q = \partial_{\theta} + i \bar\theta\, \partial_t.
\end{eqnarray} 
Application of (8) in (7) (with $A = 0$) leads to the following transformations\footnote{Basically, the 
transformations in (9) (and (10)) are {\it global} SUSY transformations because the parameters 
($\varepsilon, \bar\varepsilon$) are time-independent
for the 1D system of a {\it free} SUSY particle. One of the key characteristic features of a 
$\mathcal{N} = 2$ SUSY theory is the existence of these nilpotent (but {\it not} absolutely anticommuting)
 symmetries. (see, also, Sec. V for more discussions). } 
\begin{eqnarray}
&&\delta_1 x = i\,\bar\varepsilon\, \psi, \quad\qquad \delta_1 \psi = 0, 
\quad\qquad \delta_1 \bar\psi = - \bar\varepsilon\, \dot x, \nonumber\\
&&\delta_2 x = i\,\varepsilon\, \bar\psi,\quad \qquad \delta_2 \bar\psi = 0, 
\quad\qquad \delta_2 \psi = - \varepsilon\, \dot x.
\end{eqnarray}
Infinitesimal versions of the fermionic ($s_1^2 = s_2^2 = 0$) transformations ($s_1, s_2$) can 
be derived from the above bosonic infinitesimal
transformations by defining $\delta_1 = \bar\varepsilon\, s_1,\,\delta_2 = \varepsilon \,s_2 $. 
These fermionic ($s_1^2 = s_2^2 = 0$) transformations, for the Lagrangian (6), are
\begin{eqnarray}
s_1 x = i\, \psi, \quad\qquad s_1 \psi = 0, \quad\qquad s_1 \bar\psi = -  \dot x, \nonumber\\
s_2 x = i\, \bar\psi, \quad\qquad s_2 \bar\psi = 0, \quad\qquad s_2 \psi = -  \dot x.
\end{eqnarray}
It is elementary to check that  the action integral $S = \int dt \,L_0$ remains invariant under 
the above  SUSY transformations  $s_1$ and $ s_2$ because we have the following 
\begin{eqnarray}
s_1\, L_0 = 0, \qquad \qquad\quad s_2\, L_0 = \frac{d}{dt} (i\,\dot x\,\bar\psi).
\end{eqnarray}
Using Noether's theorem, it is straightforward to derive the conserved ($\dot Q = 0, \,\dot{\bar Q} = 0$) 
charges $Q = (i\, \dot x\,\psi)$ and $\bar Q = (i\, \dot x\,\bar\psi)$
which are nilpotent of order two (i.e. $Q^2 = \bar Q^2 = 0$).
In the forthcoming sections, we shall derive the transformations (10) by our novel 
supervariable approach.

\section{On-shell nilpotent SUSY symmetry transformations generated by Q: Anti-chiral supervariables}

The central aim of our present section and the forthcoming section is to capture the 
SUSY transformations $s_1$ and $s_2$ [cf. (10)] in the language of  Grassmannian derivatives 
($\partial_\theta,\, \partial_{\bar\theta}$) defined on the (1, 2)-dimensional supermanifold. 
Towards this goal in mind, we note that the fermionic ($s^2_1 = s^2_2 = 0$)
transformations (10) are {\it not} absolutely anticommuting (i.e. $\{s_1,\, s_2\} \ne 0$). 
As a consequence, we {\it cannot} have expansions like (1) to capture these symmetry transformations 
in the language of ordinary super derivatives ($\partial_{\theta}, \, \partial_{\bar\theta}$) w.r.t.
the Grassmannian variables $\theta$ and $\bar\theta$. In other words, we have to truncate the 
expansion (1) to derive the fermionic ($s_1^2 = s_2^2 = 0$) 
transformations $s_1$ and $s_2$ {\it independently} from the $\mathcal{N} = 2$ 
SUSY invariant restrictions (SUSYIRs) on the (anti-)chiral supervariables.

To accomplish  the above goals, first of all, we focus on the derivation of $s_1$ in the language of 
the Grassmannian derivative  $\partial_{\bar\theta}$ which is defined on the anti-chiral 
(1, 1)-dimensional super-submanifold (parametrized {\it only} by $t$ and $\bar\theta$) 
of the general (1, 2)-dimensional supermanifold. As a first step, we generalize the basic dynamical 
variables $x(t), \psi (t), \bar\psi (t) $ onto the (1, 1)-dimensional
anti-chiral super-submanifold as:
\begin{eqnarray}
&&x(t) \; \longrightarrow \; {X(t, \theta, \bar\theta)}\mid_{\theta = 0} \,= X(t, \bar\theta), \nonumber\\
 &&X(t, \bar\theta) = x(t) +  \bar\theta\, f(t),\nonumber\\
&&\psi(t) \;\longrightarrow \; \Psi (t, \theta, \bar\theta)\mid_{\theta = 0} 
= \Psi (t, \bar\theta),\nonumber\\ &&
\Psi (t, \bar\theta) = \psi (t)  + i\, \bar\theta\, b_1 (t), \nonumber\\
&&\bar\psi(t) \;\longrightarrow \; \bar\Psi (t, \theta, \bar\theta)\mid_{\theta = 0}
= \bar\Psi (t, \bar\theta),\nonumber\\ &&
\bar\Psi (t, \bar\theta) = \bar\psi (t)  + i\, \bar\theta\, b_2 (t),
\end{eqnarray} 
where the expansions of the supervariables  ($X, \Psi, \bar\Psi$) are along $\bar\theta$-direction {\it only} and 
($b_1, b_2$) are the bosonic secondary variables and $f(t)$ is a fermionic   secondary variable that would be determined by
exploiting the SUSY invariant restrictions on the anti-chiral supervariables in a specific fashion. This 
approach would be {\it physically} more appealing than the standard mathematical method  used in Sec. 2
because we shall deal with the SUSY invariance of our present theory.

We observe that the variable $\psi(t)$ is an invariant quantity under the transformation $s_1$ [cf. (10)].
We demand that such kind of SUSY invariant quantities
 {\it should not} depend on the Grassmannian variable $\bar\theta$
of the (1, 1)-dimensional anti-chiral super-submanifold. Thus, we impose  the following 
restriction on the supervariable:
\begin{eqnarray}
\Psi (t, \bar\theta) = \psi (t) \qquad\Longrightarrow \qquad b_1(t) = 0.
\end{eqnarray}
We also note that $s_1(x\,\psi) = 0$ due to the fermionic ($\psi^2 = 0$) nature of the variable $\psi(t)$.
Thus, we put the SUSYIR on the composite anti-chiral supervariables as
\begin{eqnarray}
X(t, \bar\theta)\,\Psi (t, \bar\theta) = x(t)\, \psi (t).
\end{eqnarray}
Using (13), we obtain the following relationship:
\begin{eqnarray}
f(t)\,\psi (t) = 0.
\end{eqnarray}
Similarly, we observe that $s_1(\dot {x} \,\dot\psi) = 0$ which, finally, implies that
\begin{eqnarray}
\dot{X}(t, \bar\theta)\,\dot\Psi (t, \bar\theta) = \dot{x}(t)\, {\dot\psi} (t)\quad 
\Longrightarrow \quad\dot f(t)\,\dot\psi(t) = 0.
\end{eqnarray}
The non-trivial solution of (15) and (16) is $f(t) \propto \psi (t)$. For algebraic convenience, however,  
we choose here the fermionic secondary variable $f(t) = i\,\psi(t)$.

Now, we focus on (11) and draw the most important conclusion for our purpose that 
$s_1\, L_0 = 0$.  Thus, the Lagrangian function $L_0$ is itself  an invariant quantity. 
Accordingly, it should remain independent of the Grassmannian variable
$\bar\theta$. As a consequence, we have:
\begin{eqnarray}
 \frac{1}{2}\dot {X}^2(t, \bar\theta) + i\, \bar\Psi (t, \bar\theta)\, \dot{\Psi}(t, \bar\theta)
= \frac{1}{2} \,\dot x^2(t)+ i\, \bar\psi (t)\, \dot\psi (t).
\end{eqnarray}
Substitutions from (12) and (13) into the above equation imply the following:
\begin{eqnarray}
\dot x(t)\, \dot f(t) = b_2(t)\, \dot\psi(t).
\end{eqnarray}
If we take $f(t) = i\, \psi(t)$, it is evident that $b_2(t) = i \dot x(t)$. 
Thus, it is obvious that SUSY invariant restrictions (13), (14), (16) and (17) 
lead to the determination of  secondary variables 
of expansion (12), in terms of the basic variables\footnote{We have the freedom to  
choose $f (t)  = \pm \,i \,\psi (t)$ and
$b_2 (t) = \pm \,i \,\dot x (t)$ (modulo some constant multiplicative factors). 
However, we have chosen  $f (t) = +\, i\, \psi (t), \,b_2 (t) = +\, i\, \dot x (t)$ to be 
consistent with the transformations (10) which have been derived
from the  mathematical method of superspace formalism.}, as:
 $b_1(t) = 0, \, f(t) = i\,\psi(t), \, b_2(t) = i \,\dot x(t)$.

The substitution of the above secondary variables into the expansion (12) of the 
supervariables implies the following
explicit expansions for the supervariables
\begin{eqnarray}
&&X^{(1)}(t, \bar\theta) = x(t) +  \bar\theta\, (i\,\psi) \equiv x(t) + \bar\theta \,(s_1\, x),\nonumber\\
&&\Psi^{(1)} (t, \bar\theta) = \psi (t)  + \bar\theta\,(0)
 \equiv \psi(t) + \bar\theta\, (s_1\, \psi), \nonumber\\
&&\bar\Psi^{(1)} (t, \bar\theta) = \bar\psi (t)  + \, \bar\theta\, (-\dot x) 
\equiv \bar\psi (t)  + \, \bar\theta\, (s_1\, \bar\psi),
\end{eqnarray} 
where the superscript $(1)$ denotes the expansion of the supervariables after 
the SUSY invariant restrictions that lead to the
derivation of the SUSY transformations $s_1$ under which the Lagrangian $L_0$ for 
the $\mathcal {N} = 2$ SUSY free particle remains invariant
[cf. (11)]. A close look at the above expansions (19) leads to the following mapping:
\begin{eqnarray}
\frac{\partial}{\partial \bar\theta}\,\Omega^{(1)} (t, \theta, \bar\theta)|_{\theta = 0} = s_1 \omega (t) 
\quad \Longrightarrow \quad s_1 \Longleftrightarrow \partial_{\bar\theta},
\end{eqnarray}
where, in the above, the anti-chiral generic supervariable $\Omega^{(1)} (t, \theta, \bar\theta)|_{\theta = 0}$ stands for
 $ X^{(1)}(t, \bar\theta)$, $ \Psi^{(1)}(t, \bar\theta)$ and
$ \bar\Psi^{(1)}(t, \bar\theta)$ and generic variable $\omega (t)$ corresponds to 
the basic variables $x(t), \psi(t)$ and $\bar\psi (t)$
of the Lagrangian  $L_0$ defined on the one ($0+1$)-dimensional ordinary manifold. 
It is evident that $\partial^2_{\bar\theta} = 0$.
This property obviously implies the nilpotency ($s^2_1 = 0$) of the transformations $s_1$.

The starting Lagrangian $L_0$ can be generalized onto the (1, 1)-dimensional anti-chiral 
super-submanifold (of the general (1, 2)-dimensional supermanifold) as follows 
\begin{eqnarray}
L_0 \; \Longrightarrow \; {\tilde L}^{(ac)}_0 = \frac{1}{2}\,\dot {X^{(1)}}(t, 
\bar\theta)\,\dot {X^{(1)}}(t, \bar\theta)
 + i\, {\bar\Psi}^{(1)} (t, \bar\theta)\,{\dot{\Psi}}^{(1)}(t, \bar\theta),
\end{eqnarray}
where the supervariables $ X^{(1)}(t, \bar\theta), \Psi^{(1)}(t, \bar\theta), 
\bar\Psi^{(1)}(t, \bar\theta)$  are defined in (19).
Furthermore, the SUSY invariance of $L_0$ (i.e. $s_1\, L_0 = 0$) can be captured in the  following manner
\begin{eqnarray}
\frac{\partial}{\partial \bar\theta}\, {\tilde L}^{(ac)}_0 = 0 \qquad \Longleftrightarrow \qquad s_1\, L_0 = 0.
\end{eqnarray}
Geometrically, this observation shows that the SUSY invariance of the Lagrangian ($L_0$), under symmetry transformation $s_1$, 
is equivalent to the translation of the composite supervariables of the super anti-chiral Lagrangian 
(${\tilde L}^{(ac)}_0$) along $\bar\theta$-direction of the anti-chiral 
super-submanifold such that the result is zero.
Here the superscript $(ac)$ on the super Lagrangian stands for the 
anti-chiral behavior of the super Lagrangian (21).

By exploiting the Noether theorem, it is clear that the  conserved  ($\dot Q = 0$) charge  $Q$
for the nilpotent ($s^2_1 = 0$) transformation $s_1$ is $Q  = ( i\, \dot x\,\psi) $
which turns out to be the generator of the transformations $s_1$ as is evident from the following: 
\begin{eqnarray}
 s_1\, \phi = -i\, [ \phi,\, Q]_{\pm}, \qquad\qquad \phi = x,\, \psi,\, \bar\psi,
\end{eqnarray}
where the ($\pm$) signs, as the subscripts on the square bracket,
stand for the  (anti)commutator for the generic variable $\phi$ being (fermionic)bosonic in nature.
This charge can be expressed in terms of the supervariables (19) in {\it two} different ways:
\begin{eqnarray}
Q &=& \frac{\partial}{\partial \bar\theta} \left[- i\bar\Psi^{(1)}(t, \bar\theta)\,\Psi(^{(1)}(t, \bar\theta)
\right]\equiv  \int d\bar\theta
\left[- i\bar\Psi^{(1)}(t, \bar\theta)\,\Psi^{(1)}(t, \bar\theta) \right],  \nonumber\\
Q &=&\frac{\partial}{\partial \bar\theta} \left[\dot x(t) \,X^{(1)}(t, \bar\theta)\right] 
 \equiv  \int d\bar\theta
\left[\dot x(t) \,X^{(1)}(t, \bar\theta)\right].
\end{eqnarray}
In view of the mapping (20), it is pretty obvious that (24) can be {\it also} expressed as
\begin{eqnarray}
 Q = s_1 \Big(-i\, \bar\psi(t) \,\psi(t)\Big), \qquad\qquad  Q = s_1 \Big(\dot x(t)\, x(t)\Big),
\end{eqnarray} 
where the on-shell conditions ($\ddot x = 0, \,\dot\psi = 0,\, \dot{\bar\psi} = 0$) are to
 be used. Exploiting (23), it is now clear that $s_1 Q = -i\,\{Q,\, Q\} = 0$ due to the nilpotency of the transformations $s_1$ where $s^2_1 = 0$. In the language of Grassmannian derivative $\partial_{\bar\theta}$, it is obvious that $\partial_{\bar\theta}\, Q = 0$ due to $(\partial_{\bar\theta})^2 = 0$. Thus, the nilpotency of $s_1$, $\partial_{\bar\theta}$ and $Q$ are inter-related
very beautifully.

\section{On-shell nilpotent SUSY symmetry transformations generated by $\bar Q$: Chiral supervariables}

In this section, we concentrate on the derivation of $s_2$ by exploiting the SUSY invariant restrictions 
on the chiral supervariables which are the generalizations of the basic ordinary dynamical variables
$x(t),\, \psi(t)$ and $\bar\psi(t)$ onto the (1, 1)-dimensional chiral 
super-submanifold of the general (1, 2)-dimensional supermanifold.
The generalizations and their expansions along $\theta$-direction of the (1, 1)-dimensional super-submanifold are
\begin{eqnarray}
&&x(t) \; \longrightarrow \; X(t, \theta, \bar\theta)|_{\bar\theta = 0} = X(t, \theta), \nonumber\\
&& X(t, \theta) = x(t) +  \theta\, \bar f(t),\nonumber\\
&&\psi(t) \;\longrightarrow \; \Psi (t, \theta, \bar\theta)|_{\bar\theta = 0} = \Psi (t, \theta), \nonumber\\
&&\Psi (t, \theta) = \psi (t)  + i\, \theta\, \bar b_1 (t), \nonumber\\
&&\bar\psi(t) \;\longrightarrow \; \bar\Psi (t, \theta, \bar\theta)|_{\bar\theta = 0} = \bar\Psi (t, \theta), \nonumber\\
&&\bar\Psi (t, \theta) = \bar\psi (t)  + i\, \theta\, \bar b_2 (t),
\end{eqnarray} 
where ($\bar b_1, \bar b_2 $) are the bosonic variables and $\bar f$ is a fermionic  secondary variable
on the r.h.s. of expansion (26). These secondary variables would be expressed in terms of the basic variables 
(and derivatives on them) by exploiting  theoretically important restrictions on the chiral-supervariables 
defined on the (1, 1)-dimensional chiral super-submanifold.

It is elementary to note that $s_2 \bar\psi = 0$. This implies that $\bar\psi (t)$ is an 
invariant quantity under the transformation $s_2$. Thus, we demand the $\theta$-independence 
of the supervariables $\bar\Psi (t, \theta)$ which can be mathematically expressed as:
\begin{eqnarray}
\bar\Psi(t, \theta) =   \bar\psi (t) \qquad \Longrightarrow  \qquad \bar b_2 (t) = 0.
\end{eqnarray}  
 Similarly, we note that $s_2 (x\, \bar\psi) = 0$ and $s_2 (\dot x\, \dot{\bar\psi}) = 0$ when we exploit 
the fermionic property of $\bar\psi(t)$ variable (which satisfies $\bar\psi^2 = 0$). The above observations, 
together with (27), imply the following SUSYIRs on the
composite  chiral supervariables:
\begin{eqnarray}
&&X(t, \theta)\;\bar\Psi(t, \theta) =  x(t)\; \bar\psi (t), \qquad 
\dot X(t, \theta)\;\dot{\bar\Psi}(t, \theta) =  \dot x(t)\; \dot{\bar\psi} (t).
\end{eqnarray}  
The above restrictions imply $ \bar f (t)\, \bar \psi(t)  =0, \; \dot{\bar f} (t)\, \dot{\bar \psi}(t) = 0.$
The non-trivial solution for these restrictions is $\bar f (t) \propto \bar\psi(t)$.
For the algebraic convenience, however, we choose $\bar f (t) = i\bar\psi (t)$. 
Now, we take note of the SUSY invariance 
of the following quantity: 
\begin{eqnarray}
s_2 \Big[ \frac{1}{2} \,\dot x^2(t)-  i\, \dot{\bar\psi} (t)\, \psi (t)\Big]  =0.
\end{eqnarray}
As a consequence, we demand that the SUSY invariant quantity  [contained within the square bracket of (29)],
should remain independent of the Grassmannian variable $\theta$ 
when generalized onto the (1, 1)-dimensional chiral
super-submanifold. In other words, we have the following
equality in the language of mathematical  equation:
\begin{eqnarray}
\frac{1}{2}\dot {X}^2(t, \theta) - i\, \dot{\bar\Psi} (t, \theta)\, \Psi(t, \theta)
= \frac{1}{2} \,\dot x^2(t)- i\, \dot{\bar\psi} (t)\, \psi (t).
\end{eqnarray}
The substitutions of (26), along with the relationship (27), yield the following:
\begin{eqnarray}
 \dot {\bar f}(t) \, \dot x(t) = \bar b_1(t)\, \dot{\bar\psi}(t).
\end{eqnarray}
Plugging in the value  $\bar f (t)= i\,\bar\psi (t)$, we obtain $\bar b_1 (t) = i\,\dot{x} (t)$. 
Finally, we obtain the following expansions in their full blaze of glory, namely;
\begin{eqnarray}
&&X^{(2)}(t, \theta) = x(t) +  \theta\, (i\,\bar\psi) \equiv x(t) + \theta \,(s_2\, x),\nonumber\\
&&\Psi^{(2)} (t, \theta) = \psi (t)  + \theta\,(- \dot x) \equiv \psi(t) + \theta\, (s_2\, \psi), \nonumber\\
&&\bar\Psi^{(2)} (t, \theta) = \bar\psi (t)  + \theta\, (0) \equiv \bar\psi (t)  +  \theta\, (s_2\, \bar\psi),
\end{eqnarray} 
where we have inserted the values $\bar b_1(t) = i\,\dot {x} (t), \bar f (t) = i\,\bar\psi (t), \bar b_2 (t) = 0$. 
The superscript $(2)$ on the supervariables in (32) denotes the chiral supervariables that 
have been obtained after the SUSY invariant restrictions (27), (28) and (30)
have been imposed.

A close look at (32) establishes the fact that we have already derived the transformations $s_2$ of equation (10).
Furthermore, we have obtained the following mapping:
\begin{eqnarray}
\frac{\partial}{\partial \theta}\,\Sigma^{(2)} (t, \theta,  \bar\theta)\mid_{\bar\theta =  0} = s_2 \,\sigma (t) 
\; \Longrightarrow \; s_2 \;\Longleftrightarrow  \; \partial_{\theta},
\end{eqnarray}
where $\Sigma^{(2)} (t, \theta,  \bar\theta)|_{\bar\theta =  0}$ is the generic chiral supervariable [e.g. $ X^{(2)}(t, \theta), \Psi^{(2)} (t, \theta)$ 
and $\bar\Psi^{(2)} (t, \theta)$ of (32)] and $\sigma (t)$ stands for the basic variables $x(t),\, \psi(t)$ and $\bar\psi(t)$
of the Lagrangian (6) of our present theory. We also note that the nilpotency of $\partial_\theta$ (i.e. $\partial_\theta^2 = 0$) 
implies  $s^2_2 = 0$.
Geometrically, the relation (33) shows that the SUSY transformations $s_2$ on a 1D ordinary
 generic variable $\sigma(t)$ is equivalent to the translation of the corresponding generic supervariable
$\Sigma^{(2)} (t, \theta,  \bar\theta)|_{\bar\theta =  0}$ along $\theta$-direction of the (1, 1)-dimensional chiral super-submanifold.

The Lagrangian $L_0$  can be generalized to the super chiral Lagrangian ${\tilde L}^{(c)}_0$ which can be
expressed in terms of the chiral supervariables (32) in the following fashion
\begin{eqnarray}
L_0 \; \Longrightarrow \; {\tilde L}^{(c)}_0 = \frac{1}{2}\,\dot {X^{(2)}}(t, \theta)\,\dot {X^{(2)}}(t, \theta)
 + i\, \bar\Psi^{(2)} (t, \theta)\,\dot{\Psi}^{(2)}(t, \theta),
\end{eqnarray}
The invariance of the Lagrangian $L_0$ under the transformations $s_2$ can be captured in the language of the supervariables
and Grassmannian derivative ($\partial_{\theta}$) as follows:
\begin{eqnarray}
\frac{\partial}{\partial\theta}\; \Big[{\tilde L}^{(c)}_0 \Big] = \frac{d}{dt}\Big(i\, \dot x\, \bar\psi \Big)
 \; \Longleftrightarrow \; s_2\, L_0 = \frac{d}{dt}\,\Big(i\,\dot x\, \bar\psi \Big).
\end{eqnarray}
The above expression shows the invariance of action integral $S = \int dt\, L_0$ under the nilpotent 
transformations $s_2$ which can also be expressed in terms of ${\tilde L}^{(c)}_0$ and $\partial_{\theta}$.
The geometrical interpretation of the relationship in (35) for (${\tilde L}^{(c)}_0$) can {\it also} be provided 
analogous to (${\tilde L}^{(ac)}_0$), 
as we have elaborated on, after the equation (22).

The conserved ($\dot{\bar Q} = 0$) and nilpotent ($\bar Q^2 = 0$) SUSY charge $\bar Q = (i\, \dot x\, \bar\psi)$
can be expressed in terms of the supervariables [cf. expansions in (32)]
and super derivative ($\partial_\theta$) in the following {\it two} different
forms: 
\begin{eqnarray}
&&\bar Q =\frac{\partial}{\partial \theta}\, \Big[i\,\bar\Psi^{(2)}(t, \theta)\,\Psi^{(2)}(t, \theta)\Big] 
\equiv \int d\theta\,
\Big[i\,\bar\Psi^{(2)}(t, \bar\theta)\,\Psi^{(2)}(t, \theta)\Big],  \nonumber\\
&&\bar Q =\frac{\partial}{\partial\theta}\, \Big[\dot x(t)\,X^{(2)}(t,\theta)\Big]  
\equiv \int d\theta\, \Big[\dot x(t)\,X^{(2)}(t, \theta)\Big],
\end{eqnarray}
which can be re-expressed in terms of the basic variables [$x(t),\, \psi(t),\, \bar\psi(t)$] and the transformations $s_2$
in the following manner:
\begin{eqnarray}
 \bar Q = s_2 \Big(i\, \bar\psi(t)\,\psi(t)\Big), \qquad\qquad  \bar Q = s_2 \Big(\dot {x}(t)\, x(t) \Big),
\end{eqnarray}
where we have to use the on-shell condition $\dot{\bar\psi} = 0$ for the validity 
of the second expression for $\bar Q$ in (37). The expressions in (37) demonstrate 
that $s_2\,\bar Q = -i\,\{\bar Q, \, \bar Q\} = 0$ due to the nilpotency of
$s_2$ (i.e. $s^2_2 = 0$). This observation, in turn, establishes the nilpotency of $\bar Q$ (i.e. ${\bar Q}^2 = 0$) 
which can also be expressed in the language of $\partial_\theta$ 
as it is clear that $\partial_\theta\, \bar Q = 0$ due to $\partial_\theta^2 = 0$. Thus, we note that 
the nilpotency of $s_2, \, \bar Q$ and $\partial_\theta$ are inter-related in a beautiful 
fashion within the framework of our novel approach.

\section{Specific $\mathcal{N} = 2$ SUSY algebra and its interpretation }

To derive the specific $\mathcal{N} = 2$ SUSY algebra, generated by the charges $Q, \bar Q$ and 
the Hamiltonian ($H_0$) of the theory, we modify the transformations $s_1$ and $s_2$ [cf. (10)] 
by a constant multiplicative factor in the following manner [see, e.g., footnote after (18)]:
\begin{eqnarray}
&&s_1 x = \frac{i\, \psi}{\sqrt{ 2}}, \qquad s_1 \psi = 0, 
\qquad s_1 \bar\psi = - \frac{ \dot x}{\sqrt{2}}, \nonumber\\
&&s_2 x = \frac{i\, \bar\psi}{\sqrt{2}}, \qquad s_2 \bar\psi = 0,
\qquad s_2 \psi = -  \frac{\dot x}{\sqrt{2}},
\end{eqnarray}
which lead to the derivation of the conserved ($\dot Q = \dot{\bar Q} = 0$) 
and nilpotent ($Q^2 = {\bar Q}^2 = 0$) SUSY charges as 
\begin{eqnarray}
 Q = \frac{i\, \dot x\, \psi}{\sqrt {2}},\qquad\qquad  \bar Q = \frac{ i\, \dot x\, \bar\psi}{\sqrt {2}}.
\end{eqnarray}
The canonical Hamiltonian is $H_0 = \dot x p + \dot\psi \Pi_{\psi} - L_0 = p^2/2$ where 
$p = \dot x, \Pi_{\psi} = - i \bar\psi$ are the canonical momenta w.r.t.
$x$ and $\psi$ from the Lagrangian (6).

Using the basic canonical quantum (anti)commutators
 $\{\psi,\, \bar\psi\} = -1$ and $[x, \, p] = i$ (in natural units $\hbar = c = 1$),
we observe that the operators ($Q, \bar Q, H_0$) satisfy\footnote{If we modify (8)  
[i.e. $Q = \partial_{\bar\theta} + (i/2)\,\theta \, \partial_t, \; \bar Q 
= \partial_{\theta}+ (i/2)\,\bar\theta \, \partial_t $], these operators,
too, satisfy the algebra (40) (with $H_0 = i\, \partial_t$). Similarly, if we modify (3) [i.e. ${\cal D}= \partial_{\bar\theta}
 - (i/2)\,\theta \, \partial_t, \; \bar {\cal D} = \partial_{\theta}- (i/2)\,\bar\theta \, \partial_t $], 
the set (${\cal D}, \bar{\cal D}, H_0$) 
satisfies the algebra (40) except $\{{\cal D},\, \bar {\cal D}\} = - \,H_0$. The
operator $H_0$ is the Casimir operator in the sets 
($Q,\, \bar Q,\, H_0$) and (${\cal D},\, \bar {\cal D},\, H_0)$ because 
it commutes with all the other operators.} one of 
the simplest form of the $\mathcal{N} = 2$ SUSY quantum mechanical $sl(1|1)$ algebra 
(without any central extension):
\begin{eqnarray}
&& Q^2 = {\bar Q}^2 = 0, \qquad \{Q, \, \bar Q\} = H_0, \qquad
 \big[H_0,\, Q\big]\; = \; [H_0,\, \bar Q] \;= \;0,
\end{eqnarray}
which is identical to the algebra satisfied by the celebrated  de Rham cohomological 
operators\footnote{On a compact manifold without a boundary,
a set of three operators ($ d, \delta,\Delta$) is called as the de Rham cohomological 
operators of differential geometry where $d$ is the exterior derivative,
$\delta$ is the co-exterior derivative and $\Delta$ is the Laplacian operator. 
The operators $d$ and $\delta$ are connected with
each-other by the relation $\delta = \pm * d *$ where ($*$) is the Hodge duality 
operation on the above compact manifold [17-19].}
($d,\, \delta, \, \Delta$) of differential geometry [17-19], namely;
\begin{eqnarray}
d^2 = \delta^2 = 0, \qquad \{d, \, \delta\} = \Delta, \qquad [\Delta,\, d] =  [\Delta,\, \delta] = 0,
\end{eqnarray}
where $(\delta)d$ are the (co-)exterior derivatives and $\Delta = (d + \delta)^2 \equiv \{d,\, \delta\}$
is the Laplacian operator. We note that $\Delta$ and $H_0$ are the 
Casimir operators for the algebras (41) and (40), respectively, 
because both of them commute with all the rest of the operators.

The well-known relationship $\delta = \pm *\,d\,*$ can {\it also} be captured in the 
language of symmetry properties of the Lagrangian ($L_0$) of equation (6). For instance, 
it can be seen that $L_0$ remains invariant under the following discrete symmetry transformations 
\begin{eqnarray}
x \rightarrow -\,x, \quad\qquad t \rightarrow  -\,t, \quad \psi \rightarrow 
+\, \psi, \quad\qquad \bar\psi \rightarrow -\, \bar\psi.
\end{eqnarray}
This symmetry  turns out to be the analogue of the Hodge duality ($*$) operation of differential geometry
because we observe that the following interesting relationships:
\begin{eqnarray}
s_1\, \phi = \pm\, *\,s_2\,*\, \phi,\qquad \qquad \qquad \phi = x, \psi, \bar\psi,
\end{eqnarray}
are true for the generic variable $\phi = x, \psi, \bar\psi$ of the theory 
where the analogue of ($*$) operation is 
nothing but the discrete symmetry transformations (42) and ($s_1,\, s_2$) are the continuous 
symmetry transformations (38) for the Lagrangian ($L_0$).

For a duality invariant theory (see, e.g. [20]), the ($\pm$) signs on the r.h.s. of (43)
are determined by two successive operations of the discrete
symmetry transformations on a specific variable of our theory. In this context, we observe the following:
\begin{eqnarray}
*\,(*\, x) = x, \qquad *\,(*\, \psi) = -\,\psi, \qquad *\,(*\, \bar\psi) =  - \, \psi.
\end{eqnarray}
Thus, it can be readily checked, from (43), that we have the following relationships
\begin{eqnarray}
&&s_1\, x = + *\, s_2\, *\, x, \qquad s_1\, \psi = -\, *s_2\,*\,\psi, \qquad
s_1\, \bar\psi = - *\,s_2\,*\, \bar\psi.
\end{eqnarray}
It is the dimensionality of our 1D system that allows us to have a reverse relationship amongst the
continuous ($s_1, \, s_2$) and discrete symmetry ($*$) transformations, as
\begin{eqnarray}
&&s_2\, x = - *\, s_1\, *\, x, \qquad s_2\, \psi = +\, *s_1\,*\,\psi, \qquad 
s_2\, \bar\psi = + *\,s_1\,*\, \bar\psi.
\end{eqnarray}
Thus, we have provided the physical realizations of the relationship $\delta = \pm *\,d\,*$
in the language of the interplay between the continuous and discrete symmetries of our theory
of $\mathcal{N} = 2$ SUSY quantum mechanical model.

We wrap up this section with the remark that, under the discrete  transformations (42), the conserved
charges ($Q,\, \bar Q$) and the Hamiltonian ($H_0$) transform as 
\begin{eqnarray}
&&*\, Q = - \,\bar Q,  \qquad\qquad*\, \bar Q = - \,Q, \qquad\qquad *\, H_0 =  +\, H_0, \nonumber\\
&&*\,(*\, Q) = +\, Q, \qquad *\,(*\, \bar Q) = +\,\bar Q, \qquad *\,(*\, H_0) = + \, H_0.
\end{eqnarray}
The above observations establish that the specific $\mathcal{N} = 2$ SUSY 
quantum mechanical algebra (40) remains {\it duality} 
invariant as it does {\it not} change its form under any arbitrary 
number of operations of the discrete symmetry ($*$) transformations (42).
The detailed discussions about the proof of a $\mathcal{N} = 2$ SUSY quantum 
mechanical model to be a physical example of 
Hodge theory have been performed  in our earlier work (see, e.g. [15] for details).

\section {Conclusions}

In our present endeavor, we have taken the simplest $\mathcal{N} =2$ SUSY quantum mechanical model 
of a {\it free} SUSY particle to demonstrate that the SUSY symmetries of this theory can be derived from the 
physical arguments where we demand that the SUSY invariant quantities, generalized onto the (anti-)chiral
super-submanifolds, should remain independent of the Grassmannian variables. The latter are
physically {\it not}  realized by experiments. In the old literature (see, e.g. [16]), the Grassmannian 
variables have been christened as the ``soul" coordinates because they do not physically 
manifest themselves  in nature whereas the specetime coordinates have been called as  
the ``body'' coordinates because they can be realized physically
and can be measured by appropriate physical instruments in a precise  manner. Hence, a physical quantity
should remain independent of the ``soul'' coordinates.

We have provided the geometrical meaning to the nilpotent symmetry transformations [cf. (10)] 
in the language of the nilpotency of the Grassmannian derivatives $\partial_{\bar\theta}$
and $\partial_{\theta}$. It is interesting to point out that, within the framework of our novel approach, 
we have established the inter-relationships amongst the SUSY transformations, SUSY 
charges and the Grassmannian derivatives of the (anti-)chiral super-submanifolds, 
on which, the SUSY continuous (and discrete)  symmetries are realized and interpreted geometrically.
In particular, the nilpotency property  of

(i) the SUSY transformations (10),

(ii) the translation generators along the Grassmannian 
 directions, and 

(iii) the SUSY conserved charges 

\noindent
is very deeply intertwined.

We know that $\mathcal{N} = 2$ SUSY symmetry transformations are nilpotent  of order two
but they are {\it not} absolutely anticommuting. To avoid the latter property, 
we have been theoretically compelled to choose the (anti-)chiral supervariables 
in our present endeavor. In the context of gauge theories,
we have to  have full expansions of the superfields [like  (1)] 
because (anti-)BRST symmetry transformations (corresponding to a given local gauge symmetry) 
are nilpotent as well as absolutely anticommuting. These properties 
are encoded in the similar properties (i.e. $\partial_\theta^2 = \partial_{\bar\theta}^2 = 0,\;
 \theta\,\bar\theta + \bar\theta\, \theta = 0$) obeyed by the translational genrators ($\partial_\theta, \partial_{\bar\theta}$) along the ($\theta, \bar\theta$)-directions of supermanifold on which the full 
expansions are taken into account. Thus, we have made an intelligent choice of 
the (anti-)chiral supervariables so that we could avoid the anticommutativity property   
($\partial_\theta\,\partial_{\bar\theta} 
+ \partial_{\bar\theta}\, \partial_\theta = 0$) for the ${\mathcal N} = 2$ SUSY symmetries.

In our Appendix A, we have shown the generalization of our method in deriving the 
$\mathcal {N} = 2$ SUSY symmetry transformations for the system of a SUSY harmonic oscillator.
We plan to discuss this system, in great detail, in our future endeavor where
we shall try to study the phenomenological implications of our results.
In our Appendix B, for the sake of completeness,  we have derived the off-shell nilpotent symmetries for the 
1D free SUSY particle within the framework of our supervariable approach.The on-shell nilpotent symmetries of this theory  have been discussed  in the main body of our text.

It would be a very nice future endeavor to extend our present ideas in the description of the
$\mathcal{N} = 2$ SUSY quantum mechanical models of physical interest [15,21,22] which have been
recently shown by us to be the models for the Hodge theory.  Finally,  we make a passing comment
that our supervariable approach can {\it not} be applied to $\mathcal{N} = 1$ SUSY quantum mechanical model
where the symmetry and the corresponding charge are {\it not} nilpotent but the charge obeys  the algebra.\\

\noindent
{\bf Acknowledgements} \\
Two of us (SK and AS) would like to gratefully acknowledge the financial support
from UGC and CSIR, Gov. of India, New Delhi, under their SRF-schemes.\\

\vspace {0.8 cm}
\noindent
{\bf {\large Appendix  A:  On $\mathcal{N} = 2$ continuous  symmetry transformations 
for the interacting system of a SUSY harmonic oscillator}}

\vspace {0.6 cm}
\noindent
We apply our method of derivation to one of the simplest {\it interacting} 
$\mathcal{N} = 2$ SUSY system of a 1D harmonic oscillator which is described by the 
following Lagrangian (with mass $m = 1$ and natural frequency $\omega$) (see, e.g. [22] for details)  
\[L^{(0)}_H = \frac{1}{2}\,\dot x^2 + i\, \bar\psi\,\dot\psi - \frac{1}{2}\,\omega^2\, x^2 
- \omega\, \bar\psi\,\psi, \eqno(A1)\]
where $\dot x =  (dx/dt),\, \dot\psi = (d\psi/dt)$ are the generalized ``velocities" and $x(t)$ is the bosonic variable and its  
$\mathcal{N} = 2$ SUSY counterparts are fermionic variables  $\psi(t)$ and $\bar\psi(t)$ 
(with $\psi^2 = \bar\psi^2 = 0,\, \psi\,\bar\psi + \bar\psi\,\psi = 0$) which are 
function of the evolution parameter $t$.
The above Lagrangian respects the following two continuous SUSY symmetry transformations
\[s_1 x = i\, \psi, \qquad s_1 \psi = 0, \qquad s_1 \bar\psi = -  (\dot x + i\,\omega\, x),\] 
\[s_2 x = i\, \bar\psi, \qquad s_2 \bar\psi = 0, \qquad s_2 \psi = -  (\dot x- i\,\omega\, x), \eqno(A2)\]
where the infinitesimal transformations $s_1$ and $s_2$ are on-shell 
($\dot\psi + i\, \omega\,\psi  = 0 , \dot{\bar\psi} - i\, \omega\,\bar\psi =0$)
 nilpotent  ($s^2_1 =s_2^2 = 0$) of order two. It can be checked that the 
anticommutator of $s_1$ and $s_2$ generates the time translation which 
is one of the key requirements of the 
general $\mathcal{N} = 2$ SUSY theory [defined on a 1D spacetime manifold] (see, e.g. [22]).

One can obtain the off-shell nilpotent ($s_1^2 = s_2^2 =0$) continuous 
$\mathcal{N} = 2$ SUSY transformations by linearizing the potential 
(i.e. $A^2/2 -\omega\, x\,A   = - \omega^2\,x^2/2$) by introducing an auxiliary variable 
(i.e. Lagrange multiplier) $A(t)$ as follows
\[L^{(1)}_H = \frac{1}{2}\,\dot x^2 + i\, \bar\psi\,\dot\psi -\omega\, x\,A  + \frac{1}{2} A^2 
- \omega\, \bar\psi\,\psi.\eqno(A3)\]
The following $\mathcal{N} = 2$ continuous SUSY transformations 
\[ s_1 x = i\, \psi, \qquad s_1 \psi = 0, \qquad s_1 \bar\psi = -(\dot x + i\, A),
\qquad s_1 A = -\dot \psi,\] 
 \[s_2 x = i\bar\psi,\qquad s_2 \bar\psi = 0, \qquad s_2 \psi = - (\dot x - i A),
\qquad s_2 A = + \dot {\bar\psi},\eqno (A4)\]
are the {\it symmetry} transformations for the Lagrangian (A3) because 
\[ s_1 L_H^{(1)} = \frac {d}{dt}\, \Big[-\, \omega\, x\, \psi \Big], \qquad
 s_2 L_H^{(1)} = \frac {d}{dt}\, \Big[ \bar\psi\, (i\,\dot x +  A -\,\omega\, x)\Big].\eqno(A5)\]
It is evident that the action integral $S = \int dt\,L_H^{(1)}$ remains invariant under the 
above  $\mathcal{N} = 2$  SUSY transformations (A4). We can check explicitly that 
$s^2_1 = s^2_2 = 0$ {\it without} any help from the equations of motion
 ($\ddot x  + \omega^2\, x =0,\, \dot\psi + i\,\omega\, \psi = 0,\, \dot{\bar\psi} 
- i\,\omega\, \bar\psi = 0,\, A = \omega\, x$).

 At this juncture, we shall, first of all, derive the transformations (A4) by exploiting  the SUSY invariant restrictions  
 (SUSYIRs) on the (anti-)chiral supervariables  that have been defined in (12) and (26), respectively. 
 In addition  to these, we shall include the following generalizations of the auxiliary variable $A(t)$, namely;
\[ A(t) \; \longrightarrow \; {\tilde A}(t,  \theta, \bar\theta)|_{\theta = 0}
= A (t, \bar\theta) \]
\[  A (t, \bar\theta) = A(t) + \bar\theta\, f_1(t),\]
\[ A(t) \; \longrightarrow \; {\tilde A}(t,  \theta, \bar\theta)|_{\bar \theta = 0}
= A (t, \theta)\] 
\[  A (t, \theta) = A(t) + \theta\, {\bar f}_1(t),\eqno(A6)\]
onto its (anti-)chiral SUSY counterparts ${\tilde A}(t,  \bar\theta)$ 
and ${\tilde A}(t,  \theta)$ in the equations (12) and (26), respectively. It is evident that, 
in the above expansions (A6), we have the fermionic secondary variables $f_1(t)$ and 
${\bar f}_1(t)$ because $A(t)$ is bosonic variable and the pair 
($\theta, \bar\theta$) is fermionic in nature. The derivation of the off-shell 
nilpotent symmetry transformation (A4) would, finally,
enable us to derive the on-shell nilpotent symmetry (A2), too, by the substitution  $A = \omega\, x$
which emerges as the equation of motion from (A3).

We focus {\it first} on the derivation of the transformations $s_1$ of (A4) 
by applying SUSYIRs (13), (14) and (15) which lead to the derivation of the 
secondary variables $f(t)$ and $b_1(t)$ in terms of the basic variable   
 as: $f(t) = i\,\psi(t),\, b_1(t) = 0$. Next, it is clear that $s_1(\dot x + i\, A) =0$ due to the 
 off-shell nilpotency of $s_1$ because $s_1^2 \,\bar\psi= 0$. Thus, we have the following SUSYIR 
on the (super)variables:
\[ \dot X(t,  \bar\theta) +i\, {\tilde A}(t,  \bar\theta)  = \dot x(t) + i\, A(t).\eqno(A7)\]
Plugging in $f(t) = i\,\psi(t)$ in the expansion of $X(t,  \bar\theta)$, we obtain explicitly 
$f_1(t) = - \dot\psi(t)$ if we use  the expansion  from (A6) for ${\tilde A}(t,  \bar\theta)$.
Finally, we note that we have the following SUSY invariant quantity, namely;
\[s_1\,\Big[\frac{1}{2}\,\dot x^2 + i\, \bar\psi\,\dot\psi  + \frac{1}{2} A^2 \Big] =0.\eqno(A8)\]
Thus, we have the following SUSYIR on the anti-chiral (super)variables: 
\[\frac{1}{2}\,\dot {X}(t, \bar\theta)\,\dot {X}(t, \bar\theta) 
+ i\, \dot{\bar\Psi} (t, \bar\theta)\, {\Psi}(t, \bar\theta) 
+  \frac{1}{2}\, {\tilde A}(t, \bar\theta)\,{\tilde A}(t, \bar\theta)\]
\[= \frac{1}{2} \,\dot x^2(t) + i\, \dot{\bar\psi} (t)\, \psi (t) + \frac{1}{2}\, A^2(t).\eqno(A9)\]
Substitution  of our earlier results $b_1(t) = 0,\, f(t) = i\,\psi(t)$ and $f_1(t) 
= -\,\dot\psi(t)$ in the expansions of
the supervariables $\Psi(t,\bar\theta), X(t,\bar\theta)$ and ${\tilde A}(t,\bar\theta)$, 
respectively, leads to the determination of $b_2(t)$ in terms of the dynamical 
and auxiliary variables of the Lagrangian (A3) as given below:
\[b_2 (t) = i\, [\dot x(t) + i\, A(t)]. \eqno(A10)\]
Thus, ultimately, we have the following expansions for (12) and (A6):
\[ X^{(h1)}(t, \bar\theta) = x(t) + \bar\theta\, (i\,\psi) \equiv x(t) 
+ \bar\theta\, [s_1\, x(t)],\]
\[ \Psi^{(h1)}(t, \bar\theta) = \psi(t) + \bar\theta \,(0) \equiv \psi(t) 
+ \bar\theta\, [s_1\, \psi(t)],\]
\[ \bar\Psi^{(h1)}(t, \bar\theta) = \bar\psi(t) + \bar\theta\, [-(\dot x + i A)] 
\equiv \bar\psi(t) + \bar\theta\, [s_1\, \bar\psi(t)],\]
\[ {\tilde A}^{(h1)}(t, \bar\theta) = A(t) + \bar\theta \,(-\, \dot\psi) \equiv A(t) 
+ \bar\theta \,[s_1\, A(t)],\eqno(A11)\]
where the superscript $(h1)$ denotes the expansions of the supervariables after the SUSYIRs 
(in connection with the description of the SUSY harmonic oscillator). 
A close look at (A11) demonstrates that we have already derived the SUSY 
transformations $s_1$ of equation (A4) in a subtle manner and there exists 
an explicit  mapping $s_1\leftrightarrow \partial_{\bar\theta}$.

Let us now concentrate on the derivation of $s_2$ by exploiting the SUSYIRs on the chiral 
supervariables defined in (26) and (A6). Using (27) and (29), it is evident that we obtain: 
${\bar b}_2(t) = 0,\, {\bar f}(t) = i\, {\bar\psi}(t)$. The off-shell nilpotency of the transformations $s_2$
ensures that $s_2 [\dot x(t)- i\, A(t)] = 0$. Thus, we have the following SUSYIR on the (super)variables:
\[{\dot X}(t, \theta) - i\, {\tilde A} (t,\theta) = {\dot x}(t) - i\, A(t),\eqno(A12)\]
which leads to the determination of ${\bar f}_1(t) = \dot{\bar\psi}(t)$. Finally, we observe that a modified
part of Lagrangian ${ L}^{(1)}_H$ [cf. (A3)] remains invariant under $s_2$ because we have: 
\[s_2\,\Big[\frac{1}{2}\,{\dot x}^2 - i\, \dot{\bar\psi}\,\psi + \frac{1}{2}\,A^2\Big] = 0.\eqno(A13)\]
Thus, we have the following SUSYIR on the chiral (super)variables:
\[\frac{1}{2}\,{\dot X}(t, \theta)\,{\dot X}(t, \theta) - i\, \dot{\bar\Psi} (t, \theta)\, {\Psi}(t, \theta) 
+  \frac{1}{2}\,{\tilde A}(t, \theta)\,{\tilde A}(t, \theta)\]
\[= \frac{1}{2} \,{\dot x}^2(t)- i\, \dot{\bar\psi} (t)\, \psi (t) + \frac{1}{2}\, A^2(t),\eqno(A14)\]
which leads to the determination of  ${\bar b}_1(t) = i\,[\dot x(t) - i\,A(t)]$ in terms of the
dynamical and auxiliary variables of the Lagrangian (A3).
Finally, the substitution of the values: 
\[{\bar b}_2(t) = 0,\qquad {\bar f}_1 = \dot{\bar\psi}(t),\qquad {\bar f}(t) = i\,\bar\psi(t),\qquad 
{\bar b}_1(t) = i\,[\dot{x}(t) - i\, A(t)],\eqno(A15)\]
leads to the following expansions of the chiral supervariables (26) and (A6), namely;
\[ X^{(h2)}(t, \theta) = x(t) + \theta\, (i\,\bar\psi) \equiv x(t) + \theta\, [s_2\, x(t)],\]
\[\Psi^{(h2)}(t, \theta) = \psi(t) + \theta \,[-(\dot x - i A)] \equiv \psi(t) + \theta\, [s_2\, \psi(t)],\]
\[ \bar\Psi^{(h2)}(t, \theta) = \bar\psi(t) + \theta\, (0) \equiv \bar\psi(t) 
+ \theta\, [s_2\, \bar\psi(t)],\]
\[ {\tilde A}^{(h2)}(t, \theta) = A(t) 
+ \theta \,( \dot{\bar\psi}) \equiv A(t) + \theta \,[s_2\, A(t)],\eqno(A16)\]
where the superscript $(h2)$, in the above,
 denotes the supervariables obtained after the application of SUSYIRs. It is clear, from the above
expansions (A16), that we have already derived the off-shell nilpotent SUSY transformations ($s_2$) of (A4).

We wrap up this Appendix with the remarks that the supercharges $Q$ and $\bar Q$ can 
be computed by exploiting Noether's theorem and these can be expressed in terms of 
the supervariables obtained after SUSYIRs analogous to 
(24) and (36). Similarly, the Lagrangian (A3) can be expressed in terms of (anti-)chiral 
supervariables  (A11) and (A16) and the geometrical basis for the SUSY invariance of the 
Lagrangian as well as the nilpotency of  $Q$ and $\bar Q$ could be provided within 
the framework of supervariable approach. Finally, the on-shell
nilpotent symmetries (A2) can be obtained from (A11) and (A16) if we substitute 
$A = \omega\, x$ which emerges from the Lagrangian (A3) due to the Euler-Lagrange 
equation of motion w.r.t. the auxiliary variable $A(t)$.\\

\vspace {0.6 cm}
\noindent
{\bf {\large Appendix B:  On the derivation of $\mathcal{N} = 2$ off-shell nilpotent SUSY transformations for the 1D
free SUSY particle}}

\vspace{0.8 cm}
\noindent
In the main body of our present paper, we have discussed only the derivation of $\mathcal {N} = 2$ on-shell
nilpotent SUSY symmetries for the free SUSY particle.
We can {\it also} derive the off-shell nilpotent SUSY symmetries for the same system. 
Towards this goal, we note that the modified form of the Lagrangian (6), 
with an auxiliary variable $A(t)$, namely;
\[L^{(m)}_0 = A(t)\, \dot x(t) - \frac{1}{2}\,A^2(t) + i\, \bar\psi(t)\, \dot\psi(t),\eqno(B1)\]
respects the following off-shell nilpotent $\mathcal {N} = 2$ SUSY symmetries:
\[s_1 \, x = i\,\psi, \qquad s_1\, \psi = 0, \qquad s_1\, \bar\psi = -\,A, \qquad s_1\, A = 0, \]
\[s_2 \, x = i\,\bar\psi, \qquad s_2\, \bar\psi = 0, \qquad s_2\, \psi = -\,A,\qquad s_2\, A = 0,\eqno(B2)\]
because the Lagrangian (B1) transforms as follows
\[s_1\, L^{(m)}_0 = 0, \qquad\qquad s_2\, L^{(m)}_0 = \frac{d}{dt}\, (i\, A\,\bar\psi). \eqno(B3)\]
It is elementary to note that we get back $\mathcal{N} = 2$ {\it on-shell nilpotent} 
SUSY transformation (10) from (B2)
by the substitution $A = \dot x$ which is an Euler-Lagrange equation of motion from (B1).

Taking the help of expansions in (12), (26) and (A6), we can derive the SUSY 
transformations (B2) by our supervariable approach. Let us first focus on the 
derivation of $s_1$. The SUSYIRs (13), (14) and (16) lead to the derivation of
$b_1(t) = 0$ and $f(t) = i\, \psi(t)$. Furthermore, we observe that $s_1\, A(t) = 0$ 
which shows that $A(t)$ is a SUSY invariant quantity. Thus, we have the following SUSYIR [cf. (A6)]
\[{\tilde A}(t,\bar\theta) = A(t) \;\;\Longrightarrow \;\; f_1(t) = 0. \eqno(B4)\]
The above result implies that we have already obtained the exact expressions for {\it three} secondary 
variables in the expansions (12) and (A6). These are as follows:
 \[b_1(t) = 0,\quad \qquad { f}(t) = i\,\psi(t),\quad\qquad f_1(t) = 0.\eqno(B5)\]
Finally, we note that $s_1[A(t)\,\dot x(t) + i\,\bar\psi(t)\,\dot\psi(t)]  = 0$  
which lead to the following SUSYIR on the composite supervariables, namely;
\[{\tilde A}(t,  \bar\theta)  \dot X(t,  \bar\theta) + i\,\bar\Psi(t,  
\bar\theta)\, \dot\Psi(t,  \bar\theta)
= A(t)\,\dot x(t) + i\,\bar\psi(t)\,\dot\psi(t).\eqno(B6)\]
Plugging in the values from (B5) in the expansions for $\Psi(t,  \bar\theta),\, X(t,  \bar\theta)$
and ${\tilde A}(t,  \bar\theta)$ [cf. (12), (A6)], we get the expression for 
the secondary variable $b_2(t)$, namely;
\[b_2(t) = i\, A(t). \eqno(B7)\]
Thus, ultimately, we obtain the expansions for the appropriate super expansions   (12) and (A6) as follows:
\[X^{(m1)}(t, \bar\theta) = x(t) +  \bar\theta\, [i\,\psi(t)] \equiv x(t) 
+ \bar\theta \,[s_1\, x(t)],\]
\[\Psi^{(m1)} (t, \bar\theta) = \psi (t)  + \bar\theta\,(0) \equiv \psi(t) 
+ \bar\theta\, [s_1\, \psi(t)], \]
\[\bar\Psi^{(m1)} (t, \bar\theta) = \bar\psi (t)  
+ \, \bar\theta\, [- A(t)] \equiv \bar\psi (t)  + \, \bar\theta\, [s_1\, \bar\psi(t)],\]
\[{\tilde A}^{(m1)} (t, \bar\theta) = A (t)  + \bar\theta\,(0) \equiv A(t) + \bar\theta\, [s_1\, A(t)],\eqno(B8)\] 
where the superscript ($m1$) denotes the expansions of the supervariables after 
the application of the appropriate SUSYIRs. In a subtle way, we have already 
derived the off-shell nilpotent transformations ($s_1$) of (B2).

Now we focus on the derivation of the off-shell nilpotent ($s^2_2 = 0$) 
symmetry transformations $s_2$. In this connection,
we observe that the SUSYIRs (27) and (28) lead to the determination of 
${\bar b}_2 (t)= 0,\, \bar f(t) = i\, \bar\psi(t)$
in the expansions (26). Furthermore, the SUSY invariance $s_2\, A = 0$ leads 
to the following SUSYIR [cf. (A6)]
\[{\tilde A}(t,\theta) = A(t) \;\;\Longrightarrow \;\; {\bar f}_1(t) = 0. \eqno(B9)\]
We note that the following  SUSY invariance, under the off-shell nilpotent transformations ($s_2$) 
\[s_2\,[A(t)\,\dot x(t) - i\,\dot{\bar\psi}(t)\,\psi(t)]  = 0, \eqno(B10)\]
is true. Thus, we have the following SUSYIR on the composite (super)variables:
\[{\tilde A}(t,  \theta) \, \dot X(t,  \theta) -i\,\dot{\bar\Psi}(t,  \theta)\, 
\Psi(t,  \theta)   = A(t)\dot x(t) - i\,\dot{\bar\psi}(t)\,\psi(t),\eqno(B11)\]
which leads to the determination of ${\bar b}_1(t) = i\, A(t)$.

Finally, we have the following super expansions  of the appropriate supervariables in (26) and (A6)
in the language of the transformations ($s_2$), namely;
\[X^{(m2)}(t, \theta) = x(t) +  \theta\, [i\,\bar\psi(t)] \equiv x(t) + \theta \,[s_2\, x(t)],\]
\[\Psi^{(m2)} (t, \theta) = \psi (t)  + \theta\,[- A(t)] \equiv \psi(t) + \theta\, [s_2\, \psi(t)],\]
\[\bar\Psi^{(m2)} (t, \theta) = \bar\psi (t)  + \, \theta\, (0) \equiv \bar\psi (t)  + \, \theta\, [s_2\, \bar\psi(t)],\]
\[{\tilde A}^{(m2)} (t, \theta) = A (t)  + \theta\,(0) \equiv A(t) + \theta\, [s_2\, A(t)], \eqno(B12)\] 
where the superscript ($m2$), on the supervariables, denotes the expansions obtained 
after the application of the appropriate SUSYIRs. A close look at the expansions (B8) 
and (B12) demonstrates that there exists a connection between the off-shell
nilpotent symmetries $s_1$ and  the translation generator $\partial_{\bar\theta}$ 
along the $\bar\theta$-direction of 
the anti-chiral super-submanifold [of the general (1, 2)-dimensional supermanifold]. 
In exactly similar fashion, we have the mapping: $s_2\leftrightarrow \partial_\theta$
which demonstrates the connection between the nilpotent symmetry transformations $s_2$ and
the translational generator $\partial_\theta$ along the $\theta$-direction of the 
(1, 1)-dimensional chiral super-submanifold.
We conclude that the nilpotency ($s_1^2 = s_2^2 = 0$) property of the $\mathcal{N} = 2$ 
transformations (B2) has its origin in the  nilpotency 
($\partial_\theta^2 = \partial_{\bar\theta}^2 = 0$) property  of 
the (anti-)chiral Grassmannian translational generators $\partial_{\bar\theta}$ and 
$\partial_\theta$, respectively.

\end{document}